\documentclass[12pt]{article}
\catcode`\@=11
\def\@normalsize{\@setsize\normalsize{15pt}\xiipt\@xiipt
\abovedisplayskip 14pt plus3pt minus3pt%
\belowdisplayskip \abovedisplayskip
\abovedisplayshortskip  \z@ plus3pt%
\belowdisplayshortskip  7pt plus3.5pt minus0pt}
\def\small{\@setsize\small{13.6pt}\xipt\@xipt
\abovedisplayskip 13pt plus3pt minus3pt%
\belowdisplayskip \abovedisplayskip
\abovedisplayshortskip  \z@ plus3pt%
\belowdisplayshortskip  7pt plus3.5pt minus0pt
\def\@listi{\parsep 4.5pt plus 2pt minus 1pt
            \itemsep \parsep
            \topsep 9pt plus 3pt minus 3pt}}
\def\underline#1{\relax\ifmmode\@@underline#1\else
        $\@@underline{\hbox{#1}}$\relax\fi}
\@twosidetrue \relax \catcode`@=12
\catcode`\@=11

\def\section{\@startsection{section}{1}{\z@}{3.5ex plus 1ex minus
   .2ex}{2.3ex plus .2ex}{\large\bf}}

\def\ps@headings{\def\@oddfoot{}\def\@evenfoot{}
\def\@oddhead{\hbox{}\hfill
        \makebox[.5\textwidth]{\raggedright\ignorespaces --\thepage{}--
        \hfill }}
\def\@evenhead{\@oddhead}
\def\subsectionmark##1{\markboth{##1}{}}
} \ps@headings \catcode`\@=12 \relax

\def\figcap{\section*{Figure Captions\markboth
        {FIGURECAPTIONS}{FIGURECAPTIONS}}\list
        {Fig. \arabic{enumi}:\hfill}{\settowidth\labelwidth{Fig. 999:}
        \leftmargin\labelwidth
        \advance\leftmargin\labelsep\usecounter{enumi}}}
 \relax
\def\tablecap{\section*{Table Captions\markboth
        {TABLECAPTIONS}{TABLECAPTIONS}}\list
        {Table \arabic{enumi}:\hfill}{\settowidth\labelwidth{Table 999:}
        \leftmargin\labelwidth
        \advance\leftmargin\labelsep\usecounter{enumi}}}
 \relax
\def\reflist{\section*{References\markboth
        {REFLIST}{REFLIST}}\list
        {[\arabic{enumi}]\hfill}{\settowidth\labelwidth{[999]}
        \leftmargin\labelwidth
        \advance\leftmargin\labelsep\usecounter{enumi}}}
 \relax
\catcode`\@=11
\def\marginnote#1{}
\newcount\hour
\newcount\minute
\newtoks\amorpm
\hour=\time\divide\hour by60 \minute=\time{\multiply\hour by60
\global\advance\minute by- \hour}
\edef\standardtime{{\ifnum\hour<12 \global\amorpm={am}%
    \else\global\amorpm={pm}\advance\hour by-12 \fi
    \ifnum\hour=0 \hour=12 \fi
    \number\hour:\ifnum\minute<100\fi\number\minute\the\amorpm}}
\edef\militarytime{\number\hour:\ifnum\minute<100\fi\number\minute}

\def\draftlabel#1{{\@bsphack\if@filesw {\let\thepage\relax
  \xdef\@gtempa{\write\@auxout{\string
    \newlabel{#1}{{\@currentlabel}{\thepage}}}}}\@gtempa
    \if@nobreak \ifvmode\nobreak\fi\fi\fi\@esphack}
     \gdef\@eqnlabel{#1}}
\def\@eqnlabel{}
\def\@vacuum{}
\def\draftmarginnote#1{\marginpar{\raggedright\scriptsize\tt#1}}
\def\draft{\oddsidemargin -.5truein
        \def\@oddfoot{\sl preliminary draft \hfil
        \rm\thepage\hfil\sl\today\quad\militarytime}
        \let\@evenfoot\@oddfoot \overfullrule 3pt
        \let\label=\draftlabel
        \let\marginnote=\draftmarginnote
\def\@eqnnum{(\theequation)\rlap{\kern\marginparsep\tt\@eqnlabel}%
\global\let\@eqnlabel\@vacuum}  }
\def\preprint{\twocolumn\sloppy\flushbottom\parindent 1em
        \leftmargini 2em\leftmarginv .5em\leftmarginvi .5em
        \oddsidemargin -.5in    \evensidemargin -.5in
        \columnsep 15mm \footheight 0pt
        \textwidth 250mmin      \topmargin  -.4in
        \headheight 12pt \topskip .4in
        \textheight 175mm
        \footskip 0pt
\def\@oddhead{\thepage\hfil\addtocounter{page}{1}\thepage}
        \let\@evenhead\@oddhead \def\@oddfoot{} \def\@evenfoot{}
}
\def\titlepage{\@restonecolfalse\if@twocolumn\@restonecoltrue\onecolumn
     \else \newpage \fi \thispagestyle{empty}\c@page\z@
        \def\thefootnote{\fnsymbol{footnote}} }
\def\endtitlepage{\if@restonecol\twocolumn \else  \fi
        \def\thefootnote{\arabic{footnote}}
        \setcounter{footnote}{0}}  
\catcode`@=12 \relax

\def\ps@headings{\def\@oddfoot{}\def\@evenfoot{}
\def\@oddhead{\hbox{}\hfill
        \makebox[.5\textwidth]{\raggedright\ignorespaces --\thepage{}--
        \hfill }}
\def\@evenhead{\@oddhead}
\def\subsectionmark##1{\markboth{##1}{}}
} \ps@headings \relax
\newcommand{\newc}{\newcommand}
\newc{\ra}{\rightarrow}
\newc{\lra}{\leftrightarrow}
\newc{\beq}{\begin{equation}}
\newc{\be}{\begin{equation}}
\newc{\eeq}{\end{equation}}
\newc{\ee}{\end{equation}}
\newc{\bea}{\begin{eqnarray}}
\newc{\eea}{\end{eqnarray}}

\newc{\ome}{\omega}
\newc{\ba}{\begin{eqnarray}}
 \newc{\ea}{\end{eqnarray}}

\renewcommand{\l}{\varepsilon}
\newcommand{\lb}{{\varepsilon}}
\begin{document}
\def\firstpage#1#2#3#4#5#6{
\begin{titlepage}
\nopagebreak
\title{\begin{flushright}
        \vspace*{-0.8in}
{ \normalsize  hep-ph/yymmddd\\
April 2004 \\
}
\end{flushright}
\vfill {#3}}
\author{\large #4 \\[1.0cm] #5}
\maketitle \vskip -7mm \nopagebreak
\begin{abstract}
{\noindent #6}
\end{abstract}
\vfill
\begin{flushleft}
\rule{16.1cm}{0.2mm}\\[-3mm]

\end{flushleft}
\thispagestyle{empty}
\end{titlepage}}

\def\simlt{\stackrel{<}{{}_\sim}}
\def\simgt{\stackrel{>}{{}_\sim}}
\date{}
\firstpage{3118}{IC/95/34} {\large\bf Majorana Neutrino Masses from
Anomalous $U(1)$ Symmetries} {G.K. Leontaris, J. Rizos and A. Psallidas}
{\normalsize\sl Theoretical Physics Division, The University of Ioannina,
GR-45110 Ioannina, Greece\\[2.5mm]
 }
{We explore the possibility of interpreting the solar and
atmospheric neutrino data within the context of the Minimal
Supersymmetric Standard Model augmented by a single $U(1)$
anomalous family symmetry spontaneously broken by non--zero vacuum
expectation values of a pair of singlet fields. The symmetry
retains a dimension-five operator which provides Majorana masses
for left-handed neutrino states. Assuming symmetric lepton mass
matrices, the model predicts inverse hierarchical neutrino mass
spectrum, $\theta_{13}=0$ and  large mixing  while at the same
time it provides acceptable mass matrices for the charged
fermions. }

\vskip 3truecm

\newpage

\section{Introduction}

Recent neutrino oscillation data~\cite{Fukuda:2000np,rdata} imply
that neutrino squared mass differences are tiny, with $\Delta
m^2_{\odot}\approx 10^{-5} eV^2$ and $\Delta m^2_{atm}\approx
10^{-3}eV^2$. Moreover, atmospheric neutrino mixing is rather
maximal ($\theta_{atm.}\approx \frac{\pi}{4}$), while the
corresponding solar neutrino mixing is large.  These experimental
facts suggest that the Yukawa couplings related to neutrino masses
are highly suppressed compared to those of quarks and charged
leptons while their mixing is much larger than that of the quark
sector.

Large mixing may indicate an underlying  structure of the mass
matrix determined by a symmetry beyond  the Standard Model gauge
group. A natural candidate would be an additional $U(1)$ family
symmetry that is broken at some high scale $M$, a scenario
proposed some time ago for the explanation of the charged fermion
mass hierarchy~\cite{Froggatt:1979nt, ir, mt, Sato,
Leontaris:1999wf}.
Several theoretical proposals have been put
forward in the last few years to interpret neutrino data by
additional family
symmetries~\cite{Altarelli:1998nx,Ellis:1998nk,Frampton:2004ud,
Altarelli:2004jb,Honda:2004qh,Ellis:2000js}. These attempts are
also motivated by the fact that the majority of string models
constructed so far  include several (possibly anomalous)
additional $U(1)$'s.

 It is interesting therefore to explore whether a simple extension
 of the Standard Model (SM) gauge symmetry may predict an approximate form of the
 leptonic mixing matrix. Our aim is to provide a solution using only the
minimal fermion spectrum of the supersymmetric version of the
Minimal Supersymmetric Standard Model (MSSM). Thus, we will
interpret the experimental data, without introducing the
right-handed neutrinos. Indeed, this is possible since  recent
data can be well fitted with left-handed neutrinos alone as in the
context of MSSM, there exists a single
operator\cite{Weinberg:sa,Barbieri:1980hc}\footnote{For recent
reviews see also~\cite{King:2003jb,Smirnov:2003xe}} suppressed by
one power of mass which can provide  Majorana masses for all three
neutrinos. This lepton number violating operator has the form
\ba
 \frac{\zeta_{\nu}^{a \beta} }{M }
  (\bar{L_{a}^{c}}^{i} H^{j} \epsilon_{ji})
  (H^{l} L_{\beta}^{k}  \epsilon_{lk}) &
  \equiv& \frac{\zeta_{\nu}^{a \beta}v^2 }{M }
   \bar\nu_{La}^c\nu_{L\beta}\label{neff}
  \ea
where, $\zeta_{\nu}^{a \beta}$ is an effective Yukawa coupling
depending on the details of the theory, $v$ is a Higgs vacuum
expectation value (vev) which is of the order of the electroweak
scale and $M$ stands for a large scale that will turn  out to be
of the order $10^{13-14}$ GeV. This scale is quite low to be
identified with the GUT  or the string scale in the context of the
heterotic string theory, however, it is compatible with the
effective gravity scale   in theories with large extra
 dimensions  obtained in the context of Type I string models.
 \footnote{For a similar argument, see also~\cite{Berezinsky:2004zb}}

In the present work we  explore the possibility  that neutrino masses and mixing
can be interpreted with the help of an additional anomalous $U(1)$ family symmetry
which at the same time is responsible for the generation of charge fermion mass hierarchy.
This symmetry could be anomalous and anomaly cancellation is assumed to happen
in the context of a fundamental theory valid above the scale $M$. We show that
in a generic model an additional abelian symmetry can account for atmospheric data
and predicts $\theta_{13}=0$. We also show how secondary effects possibly arising
from additional singlet(s) or some alternative mechanism, as supersymmetry breaking,
can under certain assumptions render the model compatible with all recent experimental
 data. We finally derive explicit charge assignments that reproduce the above results.

\section{\label{secc}Description of the Model}

We  consider the MSSM with gauge symmetry $G_{SM}=SU(3)\times
SU(2)_L\times U(1)_Y$ as an effective field theory below a scale
$M$ of a fundamental theory. In the context of  the $G_{SM}$
symmetry, all  gauge invariant Yukawa terms relevant to quark and
charged lepton masses appearing at the tree-level superpotential
are
\ba {\cal W} &=& y_{ij}^u Q_i
U^c_j H_2 + y_{ij}^dQ_i D^c_j H_1
               +y_{ij}^e L_{i} E^c_j H_1.
\label{sup1}
\ea
In the case of models constructed in the framework of string theory, there are
explicit examples where the MSSM fields are charged under (at least) one
additional abelian anomalous ($U(1)_X$) factor that prevents terms not invariant
under this symmetry from appearing in (\ref{sup1}). Usually, the appearance of the
additional $U(1)_X$ symmetry is accompanied by at least a pair of MSSM singlets
($\Phi$, $\bar{\Phi}$) with opposite $U(1)_X$-charges. $\Phi$ and  $\bar{\Phi}$
can acquire vevs leading to the breaking of the extra abelian symmetry.

Assuming natural values of the Yukawa couplings $\lambda_{ij}$ in
(\ref{sup1}) (i.e., order one), and taking into account the
observed low energy hierarchy of the fermion mass spectrum, we
infer that only  couplings associated with the third generation
should remain invariant at tree-level. Mass terms for the lighter
fermions are to be generated from higher order non-renormalizable
superpotential couplings. Such higher order invariants are formed
by adding to the non-invariant tree-level coupling an appropriate
number of $U(1)_X$-charged singlet fields which compensate the
excess of the $U(1)_X$-charge. In the case supersymmetric models,
the magnitudes of the singlet vevs $\langle\Phi\rangle$ and
$\langle\bar\Phi\rangle$ are related by the $D$-flatness
conditions of the superpotential, while perturbative
considerations require that the vevs for the singlet fields are
about one order of magnitude below the effective theory scale $M$
scale, therefore lighter generations couplings will be suppressed
by powers of $\lambda$, $\bar{\lambda}$ where
\ba
\lambda=\frac{\langle\Phi\rangle}{M},\,
 \bar{\lambda}=\displaystyle\frac{\langle{\bar\Phi}\rangle}{M}\label{exp}
\ea
Introducing the generic charge $U(1)_X$-charge assignments of
Table~\ref{uxa}, the charges of the entries of  the corresponding
mass matrices are
\ba
C^u_{ij}=q_i+u_j\ ,\  C^d_{ij}=q_i+d_j\ ,\
C^e_{ij}=\ell_i+e_j.
\label{cd}
\ea
Restricting  the analysis to the investigation of symmetric fermion mass matrices
\begin{table}
\begin{center}
\begin{tabular}{|cc|cc|}
\hline
Fermion&Charge&Higgs&Charge\\
\hline
$Q_i(3,2,\frac 16)$&$q_i$&$H_1(1,2,-\frac 12)$&$h_1$\\
$D^c_i(\bar 3,1,\frac 13)$&$d_i$&$H_2(1,2,\frac 12)$&$h_2$\\
$U^c_i(\bar 3,1,-\frac 23)$&$u_i$&&\\
$L_i(1,2,-\frac 12)$&$\ell_i$&$\Phi(1,1,0)$&$+1$\\
$E^c_i(1,1,1)$&$e_i$&$\bar\Phi(1,1,0)$&$-1$\\
\hline
\end{tabular}
\end{center}
\caption{\label{uxa}$U(1)_X$ charge assignments for MSSM fields.
The $U(1)_X$ charges of the two extra singlet  fields $\Phi$
and $\bar\Phi$, are taken to be $+1$ and $-1$ respectively. }
\end{table}
we obtain the following constraints
\ba
q_i+u_j&=&q_j+u_i\nonumber\\
q_i+d_j&=&q_j+d_i\label{symcon}\\
\ell_i+e_j&=&\ell_j+e_i.\nonumber
\ea
Moreover, the requirement that the third generation mass couplings appear at tree-level
imposes the additional constraints
\ba
q_3+u_3+h_2&=&0\nonumber\\
q_3+d_3+h_1&=&0\label{treecon}\\
\ell_3+e_3+h_1&=&0.\nonumber
\ea
Since in our configurations the top, bottom and $\tau$--Yukawa couplings are
equal at the high scale $M$, up to order one coefficients,  the difference
between the top mass $(m_t)$ and the bottom mass ($m_b$) must arise mainly
from a large Higgs vev ratio $\tan\beta =\frac{v_2}{v_1}\gg 1$. The case of
small $\tan\beta \sim {\cal O}(1)$  can also be worked out easily in a similar
way, by modifying conditions (\ref{treecon}) so that the $b-\tau$ Yukawa couplings
appear at a higher order.

From the above  we  conclude that the general form of the superpotential couplings
contributing to the fermion  mass matrices, are divided into two categories: \\ $a)$
The tree-level couplings
\ba
{\cal W}_{tree} &=& y_{33}^u Q_3 U^c_3 H_2 +
y_{33}^dQ_3 D^c_3 H_1
               +y_{33}^e L_{3} E^c_3 H_1
\label{sup2tree}
\ea
with $y_{u,d,e}$ being the order-one Yukawa couplings, and\\
$b)$ non-renormalizable contributions of Yukawas allowed by the
$GS\times U(1)_X$ gauge symmetry. For the up, down quarks and charged
leptons these are
\ba
{\cal W}_{n.r.}^{(1)}\propto Q_i U_j^c H_2 \varepsilon^{C^u_{ij}}+
 Q_i D_j^c H_1 \varepsilon^{C^d_{ij}}
 + L_i E_j^c H_1 \varepsilon^{C^l_{ij}} \nonumber
\ea
where, $C_{ij}^a$   ($a=u,d,l$) are defined in (\ref{cd}) and
$\varepsilon$ is defined as follows
\ba
\varepsilon^{k}=\left\{
\begin{array}{ll}
\lambda^{k}&\mbox{if}\ k=[k]<0\\
\bar{\lambda}^{k}&\mbox{if}\ k=[k]>0\\
0&\mbox{if}\ k\ne[k]
\end{array}
\right.
\label{edef}
\ea
where $[k]$ stands for the integer part of $k$. As far as neutrinos are concerned,
these are massless at tree-level, however, the non-renormalizable mass term (\ref{neff})
leads directly to a light Majorana mass matrix involving only the left handed
components $\nu_{Lj}$
\ba
  {\cal W}_{n.r.}^{(2)}= \frac{\zeta_{\nu}^{a \beta} }{M}
  \varepsilon^{C^\nu_{ij}}(\bar{L_{a}^{c}}^{i} H_{2}^{j} \epsilon_{ji})
  (H_{2}^{l} L_{\beta}^{k}  \epsilon_{lk}) &
  \equiv& {\zeta_{\nu}^{a \beta} }\,\varepsilon^{C^\nu_{ij}}\,
 \frac{v_2^2}{M}\,\bar\nu_{La}^c\,\nu_{L\beta}\label{d5}
  \ea
with $v_2=\langle{H_2}\rangle \approx O(m_W)$ and ${C^\nu_{ij}}=2h_2+\ell_i+\ell_j$.

Conditions (\ref{symcon}),(\ref{treecon})  imply that the
$U(1)_X$-charges of the up and down quark entries are equal.
Furthermore, the quark charge-entries can be written  only in terms of two
combinations, namely $q_1-q_3$ and $q_2-q_3$.  In addition to  the
conditions (\ref{symcon}),(\ref{treecon}), in order to obtain acceptable
quark mass matrices we further need to
impose~\cite{Leontaris:1999wf}
\be
q_1-q_3=\frac{n}{2}\ , \
q_2-q_3 =\frac{m}{2}\ \ \mbox{\rm where}\ m+n\ne0,\
m,n=\pm1,\pm2,\dots\label{qi}
\ee
thus, quark matrices depend only
on the two integers $m,n$.
\footnote{In the context of heterotic string theory, anomaly cancellation
conditions imply further relations between $q_i$ and $\ell_i$ \cite{ir}.
These relations impose further constraints on the $U(1)_X$
charges~\cite{Leontaris:1999wf}.}

The corresponding entries of the quark matrices take the form
\ba
C_{q}=C_{d}&=&\left(\begin{array}{ccc}
n&\frac{m+n}{2}&\frac{n}{2}\\
\frac{m+n}{2}&{m}&\frac{m}{2}\\
\frac{n}{2}&\frac{m}{2}&0\\
\end{array}\right)\label{qna}
\ea
Similarly for leptons we define the parameters $2n'=l_1-l_3$ and
$2m'=l_2-l_3$, where $m',n'$ are integers and the associated
$U(1)_X$--charge matrix takes the form
\ba
C_{e}=\left(\begin{array}{ccc}
n'&\frac{m'+n'}{2}&\frac{n'}{2}\\
\frac{m'+n'}{2}&{m'}&\frac{m'}{2}\\
\frac{n'}{2}&\frac{m'}{2}&0\\
\end{array}\right)
\label{lep1}
\ea
The zero charge in the position 33 of the above charge-matrices is due to the fact
that we demand the appearance of the corresponding Yukawa couplings at the tree-level
superpotential. For the remaining entries, a proper power of the appropriate expansion
parameter is needed.

We can re-express the generic fermion charges of Table 1 in terms of the new parameters
which we choose to be $m,n$, $m',n'$ that appear in the quark and charged lepton matrices
and $q_3,\ell_3,h_2,h_1$. The resulting assignments are presented in Table 2.
\begin{table}
\centering
\begin{tabular}{|l|c|c|c|}
\hline
field&\multicolumn{3}{c|}{generation}\\
\hline
&1&2&3\\
\hline
$Q$&$\frac{n}{2}+q_3$&$\frac{m}{2}+q_3$&$q_3$\\
$U^c$&$\frac{n}{2}-q_3-h_2$&$\frac{m}{2}-q_3-h_2$&$-h_2-q_3$\\
$D^c$&$\frac{n}{2}-q_3-h_1$&$\frac{m}{2}-q_3-h_1$&$-h_1-q_3$\\
$L$&$\frac{n'}{2}+l_3$&
$\frac{m'}{2}+l_3$&$l_3$\\
$E^c$&$\frac{n'}{2}-l_3-
h_1$&$\frac{m'}{2}-l_3-h_1$&$-h_1-l_3$\\
\hline
\multicolumn{4}{|c|}{Higgs}\\
\hline
$H_1$&$h_1$&$H_2$&$h_2$\\
\hline
\end{tabular}
\caption{Fermion $U(1)_X$ charge assignments after introducing the integer
parameters $m,n$ and $m',n'$ that appear in the quark and charge lepton matrices
respectively.}
\label{tsol1}
\end{table}

The $U(1)_X$-charge entries for  the light Majorana neutrino mass matrix take
the form
\ba
C_{\nu}= \left(\begin{array}{ccc}
n'+{\cal A}&\frac{m'+n'}{2}+{\cal A}&\frac{n'}{2}+{\cal A}\\
\frac{m'+n'}{2}+{\cal A}&{m'}+{\cal A}&\frac{m'}{2}+{\cal A}\\
\frac{n'}{2}+{\cal A}&\frac{m'}{2}+{\cal A}&{\cal A}\\
\end{array}\right)
\ea
where we have introduced the new parameter
\ba
  {\cal A}= 2 (l_3+h_2)
  \ea
We observe that the neutrino $U(1)_X$--charge entries differ from the
corresponding charged leptonic entries by the constant ${\cal A}$
\ba
  C^{\nu}_{ij} = C^{e}_{ij} + {\cal A}\label{diffNL}
\ea
The entries of the charged lepton mass matrix $C^e_{ij}$ are
integers or half-integers, therefore, in order to obtain non-zero
entries in the Majorana mass matrix too, the parameter ${\cal A}$
has to be either integer or  half-integer. If ${\cal A}$ is an
integer, an additional condition should be satisfied to insure
mixing effects in the neutrino sector. Indeed, if both
$C^{\nu}_{ij}$ and $ C^{e}_{ij}$ charge entries have the same
sign, then the corresponding mass matrix elements are proportional
by the same proportionality factor $\lb^{\cal A}$, thus both
matrices can be diagonalised simultaneously and the leptonic
mixing matrix equals to the identity. Nevertheless, in this case,
we could  obtain  mixing effects if  some of the charge matrix
elements satisfy the condition $C^{\nu}_{ij} \cdot C^{e}_{ij}\le
0$. Then, according to (\ref{edef}) the charged lepton and
neutrino mass matrices involve different expansion parameters,
thus they are no longer proportional. On the contrary, there are
no constraints if ${\cal A}$ is half integer and we will analyze
this case in the sequel as it is more promising.

\section{Neutrino Masses and Mixing}

In this section we search for explicit ${U(1)}_X$ charge assignments for MSSM particles
that provide phenomenologically acceptable mass textures for all MSSM fermions and in
particular for neutrinos. The basic structure of the mass matrices and mixing angles which
meet the phenomenological requirements can be obtained without referring to a set of particular
$U(1)_X$-charges. Explicit examples with sets of charges for all fermion and Higgs fields will be
 given in the end of this section. Before we present viable cases, we should note that our procedure
exhibits here the basic structure of the mass matrices and mixing. The most striking feature,
is that the extension of the $G_{SM}$ symmetry to include an $U(1)_X$ anomalous factor can
reproduce the correct hierarchy of all fermion fields while at the same time the recent
neutrino oscillation data are interpreted to a good approximation  by a lepton mixing matrix
involving two  mixing angles, one originating from the charged leptonic matrix matrix and the
second by the light Majorana mass matrix. However, at this level of analysis the value of the
non-vanishing coefficients of the Yukawa superpotential terms are unknown, since their
calculation requires a detailed knowledge of the fundamental theory above the scale $M$
(possibly string theory). Hence, in the present analysis, we restrict ourselves in the description
 of the general characteristics of the theory, which are nevertheless very interesting.

We first note that in our framework the quark mass matrix depends only on $m,n$ while the
leptonic one depends  on $m', n'$. We can thus fix  the parameters $m,n$, so that a correct
hierarchical quark mass spectrum is obtained.  The lepton sector can be then worked
out independently, choosing appropriate values for the two additional parameters $m'$ and $n'$.

In terms of the $\l$ parameter defined in (\ref{edef}) and up to order-one coefficients
the quark mass matrices
take the form
\ba
M_{u,d}&\sim&m_0^{u,d}\left(\begin{array}{ccc}
\lb^n&\lb^\frac{m+n}{2}&\lb^\frac{n}{2}\\
\lb^\frac{m+n}{2}&\lb^{m}&\lb^\frac{m}{2}\\
\lb^\frac{n}{2}&\lb^\frac{m}{2}&1\\
\end{array}\right)
\ea
This matrix has been work out in detail in the past
~\cite{ir,mt,Leontaris:1999wf} and it is known to ensure the
hierarchical mass structure for a variety of $m,n$ pairs.

Next, in  order to obtain a viable set of lepton mass matrices and mixing,
a systematic search shows that the charge parameters  $m',n'$ should be
$n' = \mbox{odd} \ ,\ m' = \mbox{even}$. Under this choice  the charged lepton
 mass matrix takes the form
\ba
   M_e = m_0^e\left(
  \begin{array}{ccc}
  \delta\,\lb^{n'} &  0 & 0 \\
  0  &
   \lb^{m'} & \alpha\,\lb^\frac{m'}{2}  \\
   0 &
      \alpha\,\lb^\frac{m'}{2}  & 1 \\
  \end{array}
  \right)
  \label{palep3}
\ea
where we have explicitly introduced two (out of three)  order-one parameters $a$ and
$\delta$ that account for the Yukawa couplings and renormalization effects. The lepton
mass eigenvalues are
\ba
m_e=m_0^l\,\delta\,\lb^{n'}  ,\; m_{\mu}=m_0^l(1-a^2)\lb^{m'}
,\;m_{\tau}=m_0^l(1+ a^2\lb^{m'} )
\ea
Introducing $\tan(2\phi) = 2\,a\,\lb^{\frac{m'}{2}}/(1-\lb^{m'})$
the diagonalising matrix of the charged lepton sector takes the
form
\ba
V_l(\phi) &=& \left(\begin{array}{ccc}
 1&0&0\\
 0&\cos\phi&-\sin\phi\\
 0&\sin\phi&\cos\phi\\
\end{array}\right)\,.
\ea
Turning to the neutrino sector the Majorana neutrino mass matrix takes the form
 \beq
  M_{\nu}^0 =m_0^{\nu}\,\left(
  \begin{array}{ccc}
  0 &  -\lb^{ \frac{m'+n'}{2}+{\cal A}}
  & \zeta\,\lb^{ \frac{n'}{2}+{\cal A}} \\
   -\lb^{\frac{m'+n'}{2}+{\cal A}}&
   0 &  0\\
   \zeta\,\lb^{\frac{n'}{2}+{\cal A}} &
      0&  0 \\
  \end{array}
  \right)
  \label{panet3}
  \eeq
where $\zeta$ stands for an order one coefficient. This mass
matrix can be diagonalised by a unitary matrix $V_{\nu}(\omega)$,
where $\tan\ome =\zeta\, \lb^{-m'/2}$, and can lead to bimaximal
mixing in the case that the two mass matrix elements are equal
~\cite{Frampton:2004ud}.

The neutrino mass eigenvalues are $m_{\nu_1}=-m_0^{\nu'}, m_{\nu_2}=m_0^{\nu'}$ and
$m_{\nu_3}=0$, with $m_0^{\nu'}=m_0^{\nu}\,\lb^{{\cal A} +\frac{m'+n'}{2}{}}\,\sqrt{
1+\zeta^2\lb^{-m'}}$ giving at this level for the mass square differences
\ba
\Delta m^2_{atm}=\Delta m_{23}^2={(m_0^{\nu})}^2\,{\lb^{2{\cal A} +{m'+n'}{}}}({
1+\zeta^2\lb^{-m'} })\ ,\ \Delta m^2_{\odot}=\Delta m_{12}^2=0
\label{neig}
\ea
The leptonic mixing matrix $U_l^0 = V_l^\dagger(\phi) V_{\nu}(\omega)$ is given by
\ba
U_l^0&= &\left(\begin{array}{ccc}
 -\frac{1}{\sqrt{2}}& \frac{1}{\sqrt{2}}& 0 \\
-\frac{\cos(\phi+\ome)}{\sqrt{2}}&-\frac{\cos(\phi+\ome)}{\sqrt{2}}&\sin(\phi+\ome)\\
\frac{\sin(\phi+\ome)}{\sqrt{2}}&\frac{\sin(\phi+\ome)}{\sqrt{2}}&\cos(\phi+\ome)\\
\end{array}\right)\label{bimax}
\ea
The above results exhibit a number of interesting properties of
the model, that are worth mentioning at this point. We first
observe that the model predicts an inverted neutrino mass
hierarchy, since the smallest eigenvalue corresponds to
$m_{\nu_{3}}$. We further point out that the $U(1)_X$ symmetry
implies large mixing effects in the neutrino mass matrix, in
contrast to the situation of the charged fermion sector where the
mixing is small.  Moreover, at this level of approximation,   a
$zero$-entry for the element $U_{13}$ is predicted in the mixing
matrix.  The rest of the elements are determined by two angles,
$\phi$ arising from the charged lepton mass matrix diagonalisation
and
 $\ome$ arising from the neutrino mass matrix.

In what follows,  we will show how the above  scenario is implemented,
working out specific cases with  the aim to find explicit sets of
$U(1)_X$ charges which interpret the neutrino data.  For the
specific solutions  we have set $m'$ even and $n'$ odd. Then, from
the formulae of Table (\ref{tsol1}), we find that the leptons have
fractional $U(1)_X$-charges  of the form $\frac{2 k+1}4$, with $k$
integer.
\begin{table}
\centering
\begin{tabular}{|l|c|c|c|}
\hline
\multicolumn{4}{|c|}{Solution A}\\
\hline
field&\multicolumn{3}{c|}{generation}\\
\hline
&1&2&3\\
\hline
$Q$&$4$&$2$&$0$\\
$D^c$&$2$&$0$&$-2$\\
$U^c$&$4$&$2$&$0$\\
$L$&$\frac{9}{4}$&$-\frac{1}{4}$&$-\frac{5}{4}$\\
$E^c$&$\frac{11}{4}$&$\frac{1}{4}$&$-\frac{3}{4}$\\
\hline
\multicolumn{4}{|c|}{Higgs}\\
\hline
$H_1$&$2$&$H_2$&$0$\\
\hline
\multicolumn{4}{|c|}{Singlets}\\
\hline
$\Phi$&$1$&$\bar\Phi$&$-1$\\
\hline
\end{tabular}
\begin{tabular}{|l|c|c|c|}
\hline
\multicolumn{4}{|c|}{Solution B}\\
\hline
field&\multicolumn{3}{c|}{generation}\\
\hline
&1&2&3\\
\hline
$Q$&$4$&$2$&$0$\\
$D^c$&$1$&$-1$&$-3$\\
$U^c$&$4$&$2$&$0$\\
$L$&$\frac{9}{4}$&$-\frac{1}{4}$&$-\frac{5}{4}$\\
$E^c$&$\frac{7}{4}$&$-\frac{3}{4}$&$-\frac{7}{4}$\\
\hline
\multicolumn{4}{|c|}{Higgs}\\
\hline
$H_1$&$3$&$H_2$&$0$\\
\hline
\multicolumn{4}{|c|}{Singlets}\\
\hline
$\Phi$&$1$&$\bar\Phi$&$-1$\\
\hline
\end{tabular}
\caption{Examples of $U(1)_X$ charges which lead to the neutrino mass matrix
structure discussed in the text.} \label{charges}
\end{table}

Choosing for example, the values $m=4,n=8,m'=2,n'=7, h_1 =2,h_2 =0 , A =-\frac{5}{2}$
we obtain the charge assignments of solution A of Table \ref{charges} and the following
fermion mass matrices for the quarks
\ba
 M_{u,d} \sim  m_0^{u,d}\left(\begin{array}{ccc}
\lb^8&\lb^6&\lb^4\\
\lb^6&\lb^4&\lb^2\\
\lb^4&\lb^2&1\\
\end{array}\right).
\ea
(which is the texture discussed in~\cite{Leontaris:1999wf}),
the charged leptons
\ba
M_e \sim m_0^e\left(\begin{array}{ccc}
\delta\,\lb^{7}&0&0\\
0&\lb^2&a\,\lb\\
0&a\,\lb&1\\
\end{array}\right)\label{lex}
\ea
and the neutrinos
\ba
M_\nu^0 \sim m_0^\nu\left(\begin{array}{ccc}
0&-\lb^{2}&\zeta\,\lb\\
-\lb^{2}&0&0\\
\zeta\,\lb&0&0\\
\end{array}\right)\label{lexx}.
\ea

Charged lepton masses can be fit  within a range of the mass
matrix  parameters in (\ref{lex}). For example,  choosing
$\lb\sim0.28$, $\alpha\sim-1.3$ and $\delta\sim2$, the correct
mass spectrum is obtained. Atmospheric neutrino oscillation
mass-squared difference is then reproduced for $M\sim 5\times
10^{13} GeV$ modulo order one coefficients. This scale is quite
low to be identified with the string scale in heterotic
constructions, it is however compatible with type I superstring
models where the string scale is tight to the Planck scale. Other
configurations of additional $U(1)_X$-charges are also possible
since the mass matrices under consideration do not depend on the
parameters $q_3,\ell_3$. For example choosing solution B of Table
\ref{charges} we obtain the same mass matrices as in solution A
considered above.

As already noted however,  at this level of analysis,  the
neutrino mass splitting between the first and second generation
does not appear because the two eigenstates are degenerate.
Moreover, the solar neutrino mixing angle is maximal, a situation
disfavored by recent data. This discrepancy can be lifted however,
if additional non-zero entries are generated by hierarchically
smaller effects.  For example, if we assume an additional pair of
singlet fields $\chi,\bar\chi$ with $U(1)_X$-charges $\pm3/2$\, we
obtain $M_{\nu_{23}}=M_{\nu_{32}}\propto \eta$ and
$M_{\nu_{11}}\propto \bar\eta^3$ (the appearance of several
singlet fields is a usual phenomenon in string models). It is also
possible to generate the required entries by some other mechanism,
for example supersymmetry breaking.  We find it interesting that
two additional entries, for example 11 and 23,  smaller than the
entries 12 and 13 already present at this level, would be
sufficient to bring the final form of the neutrino matrix to an
acceptable two-zero texture mass matrix ~\cite{Frampton:2002qc},
that provides the necessary mass splitting  and interpret
accurately the experimental data. To show that this is indeed the
case, let us assume that, after the inclusion of these effects and
in the basis where the charged lepton mass matrix is diagonal, the
neutrino mass matrix takes the form
\ba
  M_{\nu} =m_0^{\nu}\,\left(
  \begin{array}{ccc}
  2 x &  -\cos\,\bar\ome & \sin\,\bar\ome  \\
   -\cos\,\bar\ome& 0 &  2 y\\
   \sin\,\bar\ome &  2 y&  0 \\
  \end{array}
  \right)
  \label{mncor}
  \ea
  where $\bar\omega =\ome+\phi$.
Using the above stated assumption that $x,y <
\cos\,\bar\ome,\,\sin\,\bar\ome$ the eigenvalues of (\ref{mncor})
are
\ba
m_{\nu_1}&\approx& m_0^\nu \left(-1+
x-y\sin\,(2\bar\ome)\right)\label{m1c}\\
m_{\nu_2}&\approx& m_0^\nu \left(1+
x-y\sin\,(2\bar\ome)\right)\label{m2c}\\
m_{\nu_3}&\approx&2\,m_0^\nu\,y\,\sin\,(2\bar\ome)\label{m3c}
\ea
where higher order corrections ${\cal O}(x^2,xy,y^2)$ etc, are
omitted. In terms of the above eigenvalues, the mass-squared
differences are
\ba
\Delta m_{12}^2&\approx& 4 m_0^2
(x-y\sin (2\bar\ome))\label{dm122}\\
\Delta m_{23}^2&\approx&
m_0^2\left(1-2(x-y\sin(2\bar\ome))\right)\label{dm232}
\ea
 The diagonalising matrix is then
\ba
U_l&\approx&\left(\begin{array}{ccc}
   \frac{1}{{\sqrt{2}}} +
    \frac{ x  +
         y\,\sin (2\,\bar\omega )
          }{2\,
       {\sqrt{2}}}
    &
   \frac{1}{{\sqrt{2}}} -
    \frac{x  +
         y\sin (2\,\bar\omega )
         }{2\,
       {\sqrt{2}}}
    &
   2y\cos (2\,\bar\omega ) \\
   - \frac{\cos (\bar\omega )}
       {{\sqrt{2}}}   +
    g(x,y)
    &
   \frac{\cos (\bar\omega )}
     {{\sqrt{2}}} +
    g(x,y)
    &
   \sin (\bar\omega ) \\
  \frac{\sin (\bar\omega )}
     {{\sqrt{2}}} -f(x,y)
     &
   -\frac{\sin (\bar\omega )}
       {{\sqrt{2}}} - f(x,y)
       &
   \cos (\bar\omega )
  \end{array}
  \right)\label{Ucor}
  \ea
where again we have omitted  higher order corrections and  $f(x,y), g(x,y)$
in the matrix entries stand for
\begin{eqnarray}
f(x,y)&=&\frac{\left( 5\,y \,
          \cos (\bar\omega ) +
         3\,y \,
          \cos (3\,\bar\omega ) +
         2\,x\,
          \sin (\bar\omega ) \right) }{4\,{\sqrt{2}}}\\
g(x,y)&=&=\frac{\left( 2\,x \,
          \cos (\bar\omega ) +
         5\,y \,
          \sin (\bar\omega ) -
         3\,y \,
          \sin (3\,\bar\omega ) \right)
         }{4\,{\sqrt{2}}}
\end{eqnarray}
Identifying the entries of the diagonalising matrix (\ref{Ucor})
with the standard parametrization angles $\theta_{ij}$
($\sin\theta_{ij}\equiv s_{ij}, \cos\theta_{ij}\equiv c_{ij}$),
\ba
U_{\nu}&=&\left(\begin{array}{ccc}
c_{12}c_{13}&s_{12}c_{13}&s_{13}\,e^{i\delta}\\
-s_{12}c_{23}-c_{12}s_{23}s_{13}&c_{12}c_{23}-s_{12}s_{23}s_{13}&s_{23}c_{13}\\
s_{12}s_{23}-c_{12}c_{23}s_{13}&-c_{12}s_{23}-s_{12}c_{23}s_{13}&c_{23}c_{13}
\end{array}
\right)
\ea
up to order ${\cal O}(x^2,xy,y^2)$ corrections (assuming the CP-phase $\delta=0$) we find
\ba
\tan\theta_{23}&\approx&\tan\bar\omega\\
\tan\theta_{13}&\approx&2y\cos(2\bar\omega) \\
\tan\theta_{12}&\approx&1-(x+y \sin(2\bar\omega))\label{tanq12}
\ea
Here, $\tan\theta_{23}$  differs  from the original
$\tan(\bar\ome)$ only up to second order corrections on the
parameters $x,y$, while $\tan\theta_{12}$'s value   depends on the
linear combination $(x+y \sin(2\bar\omega))$ of  $x$ and $y$. On
the other hand, the neutrino mass-squared differences have a ratio
which depends on a different $x,y$ linear combination,
$(x-y\sin(2\bar\ome)$,
\ba
\frac{\Delta m_{12}^2}{\Delta m_{23}^2}&=&\frac{4
(x-y\sin(2\bar\ome))}{1-2 (x-y\sin(2\bar\ome))}\label{ratio}
\ea
We note that it is crucial that two different combinations of $x,y$ enter in
the expressions (\ref{tanq12}) and (\ref{ratio}). Clearly, if either of $x,y$
is taken zero, the data cannot be reconciled.

The  allowed  ranges for mixing angles at $3\sigma$ are given by,
\ba
0.29\le \tan^2\theta_{12}\le 0.64\,, \\
0.31\le\sin^2\theta_{23}\le
0.72\,,\\
 \sin^2\theta_{13}<0.054\ \ \nonumber
\ea
and the mass-squared differences have the range
\ba
5.4\times10^{-5}\le \Delta\,m_{12}^2/eV^2\le 9.5\times
10^{-5},\;\;1.4\times 10^{-3}\le \Delta\,m_{23}^2/eV^2\le 3.7\times
10^{-3}\nonumber
\ea
Using relations (\ref{tanq12}), (\ref{ratio}) and  the experimental data we find that
experimentally acceptable $\tan\theta_{12}$ values can be satisfied for  $x\approx
[0.10-0.24]$ and $y\approx [0.10-0.22]$,  assuming $\bar\omega$ to be maximal.
We remark that these values in a wide portion of the acceptable range, are sufficiently
smaller that the order one 12- and 13-neutrino mass matrix entries and thus our approximation
is consistent. In this case we also have $\theta_{13}=0$.  Departing slightly from $\bar\omega=\pi/4$
we still have acceptable values for $x,y$ and $\theta_{13}$ within the experimental limits,
however the parameter space consistent with our approximation is reduced. These facts justify the
assumption that the 11, 23 elements can be generated perturbatively. We finally check the impact
of the above on the parameter related to $\beta\beta_{o\nu}$-decay effective neutrino mass.
Ignoring the tiny $m_{\nu_3}$ contribution and second order effects, this is given by
\ba
|\langle m_{ee}\rangle |\approx  2\,m_0\,y\,\sin2\bar\omega
\ea
which is one order of magnitude below the current experimental bound for the parameter region
where our perturbative approach is valid.

\section{Conclusions}

In this letter, we have presented a simple extension of the Minimal Supersymmetric
Standard model by an anomalous $U(1)_X$ symmetry broken at some high scale $M$ and
attempted to interpret the recent neutrino experimental data using just the left-handed
neutrino components. Assuming symmetric mass matrices and that the third generation of
up, down quarks and charged fermions acquire masses at tree-level, we derive the general
charge assignments for MSSM fermions and examine their implications for the Majoranna neutrino
mass matrix resulting from the dimension 5 operator $(LH)^2/M$. We find that the model leads
naturally to inverted mass hierarchy for neutrinos, $\theta_{13}=0$ and maximal atmospheric
mixing for  $M\sim10^{13-14}GeV$. At this level the absolute masses of the lightest eigenstates
are equal and solar neutrino mixing turns out to be also maximal. We show that higher order
corrections which may arise from supersymmetry breaking or additional singlet fields, lift the
mass degeneracy and the solar neutrino data can be accurately described. We derive explicit
fermion ${U(1)}_X$ charge assignments that realize the above scenario.

\newpage

{\bf Ackmowledgements}. {\it This work is partially supported by
the Hellenic  General Secretariat for Research and Technology (
research proposal ``$HPAK\Lambda{EITO}\Sigma$'').}

\end{document}